\documentclass[12pt,a4paper]{article}

\usepackage{feynmp,amsbsy,latexsym,cite}

\newcommand{\ket}[1]{\vert #1 \rangle}
\newcommand{\xb}{{\boldsymbol x}}

\newcommand{\psias}{\psi^\as}

\newcommand{\soft}{{\mathrm{soft}}}

\newcommand{\dd}{{d^{\dag}}}
\newcommand{\ad}{{a^{\dag}}}
\newcommand{\cd}{\cdot}
\newcommand{\intx}{\int\!d^3x\;}

\newcommand{\intp}{\int\!\frac{d^3p}{(2\pi)^3}\;}

\newcommand{\intkstwo}{\int\limits_{\soft}\!\frac{d^3k}{(2\pi)^32\omega_k}\;}

\newcommand{\psidirac}{\psi_D}
\newcommand{\ecd}{{\cdot}}

\newcommand{\as}{{\mathrm{as}}}
\newcommand{\ex}{{\mathrm{e}}}
\newcommand{\pa}{\partial}
\begin{document}

\begin{titlepage}
%
%
%
\begin{center}{\Large{\textbf{Charged Particles: A Builder's Guide}}\footnote{Talk held by M.~Lavelle
at the Sixth Workshop on Non-Perturbative Quantum Chromodynamics,
Paris, June 2001. To appear in the proceedings.}}\\ [12truemm]
\textsc{Emili Bagan}\footnote{email: bagan@ifae.es}\\
[5truemm] \textit{Dept.~Fisica Te\`orica \&\ IFAE\\
Edifici Cn\\
Universitat Aut\`onoma de Barcelona\\E-08193 Bellaterra  (Barcelona)\\
Spain\\[5truemm]}
\textsc{Robin Horan\footnote{email: rhoran@plymouth.ac.uk}, Martin
Lavelle}\footnote{email: mlavelle@plymouth.ac.uk} and
\textsc{David McMullan}\footnote{email: dmcmullan@plymouth.ac.uk}\\
[5truemm] \textit{School of Mathematics and Statistics\\ The
University of Plymouth\\ Plymouth, PL4 8AA\\ UK}
\\[5truemm]
\textsc{Shogo Tanimura}\footnote{email:
tanimura@kues.kyoto-u.ac.jp}\\[5truemm]
Department of Engineering Physics and Mechanics\\
Kyoto University\\
Kyoto 606-8501\\
Japan
\end{center}

\bigskip\bigskip\bigskip
\begin{quote}
\textbf{Abstract:} It is sometimes claimed that one cannot
describe charged particles in gauge theories. We identify the
root of the problem and present an explicit construction of
charged particles. This is shown to have good perturbative
properties and, asymptotically before and after scattering, to
recover particle modes.
\end{quote}

\end{titlepage}

\setlength{\parskip}{1.5ex plus 0.5ex minus 0.5ex}

\section*{Introduction}

The title may need some explanation -- after all it might seem obvious
what a charged particle is. After all a free fermion field has a plane wave expansion in terms of particle
creation and annihilation operators.  But the fermion in QED is an interacting field and this
means that it is \emph{not} easy to describe charged particles, as the following
quote from a classic paper\cite{Kulish:1970ut} on the infra-red
structure of QED makes clear: \lq\lq \dots\ \textsl{the relativistic concept of a
charged particle does not exist}\rq\rq! The aim of this talk is to explain  the problem
that led to this statement (and similar ones) and to explicitly demonstrate how it can be solved.

\section*{The Problem with Particles}

Particle descriptions in field theory are based around the creation and annihilation operators
in the plane wave expansion of free fields. In the standard LSZ approach to scattering, the assumption is made
that the coupling constant may in some sense be \lq switched off\rq\ at large times before and after scattering.
This is fine for some theories and toy models but, unfortunately, not for our paradigm unbroken gauge theories
QED and QCD. The observation of hadrons, rather than quarks or gluons, in detectors
shows this for QCD. For QED the masslessness of the photon implies a long range interaction
which only falls off as $1/r$. This is too slow to be neglected and generates the infra-red problem (see below).

Kulish and Faddeev\cite{Kulish:1970ut} found the non-vanishing form of the asymptotic
interaction Hamiltonian in QED\footnote{This work has since been extended to theories with four point
interactions\cite{Horan:1999ba}.}. They then used this to calculate the form of the Heisenberg fields
at asymptotically large times, $t$.
Since there is a residual interaction, the fields tended not to a plane wave but rather to
\begin{equation}
\psias(x)=\intp\frac1{\sqrt{2E_{\smash{p}}}}D(p,t)
\left\{
b(p,s)u^s(p)e^{-ip\ecd x}+
\dd(p,s)v^s(p)e^{ip\ecd x}
\right\}
\end{equation}
where $D(p,t)$ is an ($A_\mu$) field dependent
\textsf{d}\textmd{i}\emph{s}t\textsl{o}\texttt{r}\textit{t}\textsc{i}\textsf{o}\texttt{n} factor.
They then concluded that these distorted plane waves imply that it is impossible to describe charged
particles in unbroken gauge theories.

This conclusion is, though, not forced upon us. Rather, since the coupling does \textit{not}
vanish asymptotically, the matter field in the Lagrangian is not gauge invariant, $\psi\to \exp(-ie\theta)\psi$,
and may not be identified with a physical particle. Indeed it may be seen that the asymptotic form of the
vector potential is such that the commutator of particle creation and annihilation operators produces the
correct electro-magnetic fields for a particle with the appropriate momentum. It is therefore necessary to find the
(gauge invariant!) fields which do asymptotically tend to the particle creation/annihilation operators.

\section*{Starting to Identify the Solution}

Many years ago Dirac\cite{Dirac:1955ca} proposed
that the combination, $\mathrm{e}^{-ie\frac{\partial_iA_i}{\nabla^2}}\psi$,
should be used to describe charged particles like the electron. This was for two reasons: i) it is locally
gauge invariant, and ii) it describes the matter field surrounded by a Coulombic field:
\begin{equation}
E_i(x_0,{\boldsymbol{x}})\psidirac(y)\ket0=
-\frac {e}{4\pi}
\frac{{\boldsymbol{x}_i-\boldsymbol{y}_i}}{\vert{\boldsymbol{x}-
\boldsymbol{y}}\vert^3}\psidirac(y)\ket0\,.
\end{equation}
The only ingredients needed to obtain this result are the fundamental equal-time commutator and the standard
three dimensional Green's function for $1/\nabla^2$. We would say that this is a description of a matter field
\textit{dressed} by an electro-magnetic field. This idea has been taken up by various authors, see, e.g.,
\cite{Lavelle:1997ty,Horan:1998im,haller:1996} and references therein.
Such gauge invariant fields clearly have a chance of describing physical particles, but how do we
know which ones to use?

\section*{A Systematic Approach}

The extension to moving charges and, especially, QCD where our understanding is less firm, requires a more
systematic approach to finding the correct \emph{dressing}. Writing our dressed field as $h^{-1}\psi$,
we have two requirements\cite{Bagan:1998kg}
on the dressing, $h^{-1}$ describing a particle moving with four-velocity $u$:\\[2mm]
\begin{center}
\begin{minipage}{.97\linewidth}\begin{center}\fbox{\begin{minipage}{.87\linewidth}
\begin{center}
Local gauge invariance of $h^{-1}\psi$: which implies $h^{-1}\to h^{-1}U$\\[3mm]
An additional  \emph{dressing equation}: $u\cdot \partial h^{-1}=-ieh^{-1}u\cdot A$\\[1mm]
\end{center}
\end{minipage}}\end{center}
\end{minipage}\\[3mm]\end{center}
The first requirement is a bare minimum -- physical variables are gauge invariant -- but the latter
is new. It can be motivated  by studying the form of the asymptotic interaction Hamiltonian
in QED, and demanding that it should vanish at one point on the mass shell.
Also, if our gauge invariant, dressed matter is to have a sharp momentum, we must demand
$u\cdot \partial (h^{-1}\Phi)=0$, and in the heavy effective
theory the equation of motion for the matter field has the form $u\cdot D\Phi=0$. Combining
these last two equations immediately gives the dressing equation.

We stress again that neither the dressing nor the matter field are physical on their own. \emph{Only
the combination $h^{-1}\psi$ is locally gauge invariant and can be identified with a charged particle.}

In QED we can solve these requirements and so obtain a dressing with a rich structure. The detailed
form of the dressing can be found elsewhere, but in the static limit it becomes:
\begin{equation}
  h^{-1}(x)=\exp\left(i\ex\int_{-\infty}^{x^0}\!ds
  \frac{\pa^iF_{i0}}{\nabla^2}(s,\xb)\right)
  \exp\left(-i\ex\frac{\pa_i A_i}{\nabla^2}\right)\,,
\end{equation}
where we recognise the Dirac dressing and an additional, and separately gauge invariant structure.
We call the Dirac term, and its generalisations to a moving charge, the minimal or \emph{soft} dressing
and, for reasons explained below, we call the additional structure the \emph{phase} dressing.
The minimal term is needed for gauge invariance and, as will be explained in D.~McMullan's talk, the additional
structure contains screening effects in QED.

\section*{Perturbative Tests}
The variables which we are proposing are necessarily non-local and non-covariant. It is thus
natural to ask whether
they can be used in practical work. The answer is yes, as we have shown
in a detailed series of calculations\cite{Bagan:1998kg,Bagan:1999jk}.
For the purposes of showing the particle nature of our dressed
fields the main point to note is that the on-shell Green's functions of these gauge invariant variables have a
good pole structure. Recall that the usual on-shell Green's functions of QED, such as the fermion propagator,
and the $S$-matrix elements have IR divergences which are of two forms: soft divergences and (imaginary) phase
divergences. We have shown that this rich structure meets its match in the structure of the dressing: the soft
dressing introduces new Feynman diagrams which contain and cancel the usual soft divergences, while the additional
phase dressing removes the usual phase divergences. We thus end up with IR finite on-shell Green's functions. We stress
that this cancellation only takes place if the velocity parameter in the dressing and the velocity of the point of the
mass shell where the renormalisation takes place match. In general this requires different dressings for
different legs.
This good pole structure is of course a necessity if we want to be able to describe
particles\footnote{It should also be noted that this
good IR behaviour is not accompanied by poor UV properties: we have also seen that these fields can be
multiplicatively
renormalised and that their composite operator behaviour is good (they do not mix\cite{Bagan:1999jk}).}.

\section*{Charged Particles}
We have seen that these variables have good perturbative properties. Now we want to return to the initial
problem as raised by
Kulish and Faddeev. This was, we recall,
that the Lagrangian fermion does not asymptotically tend to a plane wave. Normally
we would, e.g., define the large time limit
\begin{equation}
  b(q,s, t):=\intx\frac1{\sqrt{2E_{\smash{q}}}}u^{\dag s}(q)\psi(x)\ex^{iq\ecd
  x}\,,
\end{equation}
to extract a particle annihilation operator. But as those authors showed, one so obtains at large
times in QED up to order $e$:
\begin{equation}
  b(q,s,t)=\left\{1-e\!\!\intkstwo\!
  \left(
  \frac{q\cd a(k)}{q\cd k}\ex^{-it\frac{k\ecd q}{E_q}} -
\frac{q\cd \ad(k)}{q\cd k}\ex^{it\frac{k\ecd q}{E_q}}
  \right)\right\}b(q,s)\,,
\end{equation}
which shows the distortion\footnote{Actually this exponentiates and there is a
further factor from the phase which we do not mention here}. However,
repeating this calculation with dressed matter yields a further structure from the soft dressing. Essentially the
factors of $\frac{q^\mu}{q\ecd k}$ dotting into the vector potential
creation and annihilation operators in the above equation are replaced by
\begin{equation}
  \frac{q^\mu}{q\ecd k}\to\frac{q^\mu}{q\ecd k}-\frac{V^\mu}{V\ecd k}
  \,,
\end{equation}
where $V^\mu=(\eta+v)^\mu(\eta-v)\cd k-k^\mu$, with $u^\mu=\gamma(\eta+v)^\mu$ being
the four velocity of the charged particle. It is now fairly easy to
see\cite{Bagan:2000mk,Bagan:1999jf} that this implies
that the distortion vanishes at the correct point on the
mass shell. This means that \emph{we have a particle interpretation!}

\section*{Summary}

Frivolously we may conclude that we are indeed entitled to call ourselves
particle physicists. More seriously we note the following:
\begin{itemize}
  \item The residual interaction means that the coupling in QED does not asymptotically vanish and so
  the matter field \emph{on its own} is never physical.
  \item Trying to ignore this fact generates the IR problem.
  \item Taking this interaction seriously implies:
\begin{itemize}
  \item we must construct \emph{gauge invariant} (dressed) descriptions of charged particles!
\end{itemize}
 \item Gauge invariance is not enough: we require a \emph{dressing equation} to describe charges
 with sharp momenta.
 \item The solutions of these two demands have gauge invariant, Green's functions which are  IR finite on-shell.
 \item They have other good perturbative properties and we have seen that they have a particle description.
\end{itemize}
The extension of this programme to QCD is sketched in D.~McMullan's contribution to these proceedings,
where it is shown that the analogue of these variables give an insight into the physics of
screening, anti-screening and indeed confinement.

\section*{Acknowledgments}
This work was supported by a
British Council/Spanish Education Ministry \textit{Acciones
Integradas} grant, a Royal Society Joint Project grant and a PPARC
Theory Travel Fund grant. We thank
Kurt Haller for a discussion during this meeting.

\end{document}